\documentclass[twoside]{article}
\usepackage{fleqn,espcrc2}

\newcommand{\AmS}{{\protect\the\textfont2
  A\kern-.1667em\lower.5ex\hbox{M}\kern-.125emS}}

\title{On the local symmetries of gravity and supergravity models}

\author{Igor A. Bandos\address[Valencia]
{Departamento de F\'{\i}sica Te\'orica and IFIC 
(CSIC-UVEG), 
 46100-Burjassot (Valencia), Spain}
\address[KIPT]{Institute for Theoretical Physics, NSC KIPT, 
UA61108,
Kharkov, Ukraine}, Jos\'e A. de 
Azc\'arraga\addressmark[Valencia] and  
Jos\'e M. Izquierdo\address[Vayad]{Departamento de F\'{\i}sica Te\'orica,
 Facultad de Ciencias, 47011-Valladolid, Spain}}

\begin{document}

\begin{abstract}
We present here a detailed analysis of the local symmetries of supergravity 
in an arbitrary dimension $D$, both in the component and superfield 
approaches, using a field--space democracy point of view. As an application, 
we discuss briefly how a complete description of the 
local gauge symmetries clarifies the properties  of 
the supergravity--superbrane coupled system in the standard
background superfield approximation for supergravity.\\
\centerline{[10th January, 2002; FUTV/02-1001, IFIC/02-02]}
\centerline{{\sl To appear in the Proc. of the} 
{\it XVIth Max Born symposium} {\sl on}}
\centerline{ {\it Supersymmetry and Quantum Symmetries}, 
{\sl Dubna-Wroclaw, 21-25 September, 2001}}

\vspace{1pc}
\end{abstract}

\maketitle

\section{Introduction}

A recent supersymmetric analysis of the 
supergravity--superbrane interaction \cite{BAIL,BAI+} in which supergravity is 
described by its group manifold action (not as a background), as well as 
other related models  \cite{Bagger,BKPO,Bergshoeff},  
makes desirable  a reexamination of the r\^ole of local supersymmetry and, more 
generally, of the gauge symmetries  in supergravity models. 

We present here a detailed 
account of local symmetries, beginning in Sec.2 by the differential forms 
formulation of $D$--dimensional general relativity. 
We  describe in Sec.3 the complete set of its 
spacetime
gauge symmetries, including diffeomorphism invariance 
whose discussion, together with general coordinate transformations, 
becomes especially relevant 
when the interacting system of (super)gravity and (super--$p$--)brane  is 
considered. 
This is so because in the super--$p$--brane action the spacetime variables 
play a dynamical r\^ole. 

We explain how the 
presence of diffeomorphism invariance provides the possibility of presenting 
the general coordinate invariance in an equivalent form, its `variational 
version',  which does not act on the spacetime coordinates 
(see \cite{WZ78} for 
the $D=4$ $N=1$ superfield supergravity case). 
In Sec. 4 we consider the second Noether theorem for  
general relativity. Then, in Sec. 5, we describe the general structure 
of the action for the standard, component formulation of supergravity, its 
local symmetries and their associated Noether identities.

Sec. 6 is devoted to the superspace general coordinate symmetry 
and other gauge symmetries for 
the on--shell superfield 
formulation of supergravity, where supergravity  
is described by the set of 
constraints on the superspace torsion, which imply dynamical equations. 
We apply this knowledge in Sec.7 to uncover the relation between the local 
supersymmetry and the $\kappa$--symmetry of the  supermembrane in 
a $D=11$ superfield 
supergravity background.

\section{D--dimensional General Relativity in differential form}

$D$--dimensional gravity models can be formulated in terms of 
the moving frame or vielbein fields  
$e_{\mu}^{~ {a}}(x)$ (tetrad in $D=4$), 
which determine the vielbein one--forms on spacetime $M^D$,  
\begin{eqnarray}\label{ge}
& \qquad e^a(x)= d{x}^{\mu} e_{\mu}^{~ {a}}(x)\; .  
\end{eqnarray} 
A  change of frame is given by a matrix of the local 
$SO(1,D-1)$ group, 
\begin{eqnarray}
 \label{LL0} 
& e^{a}(x) \rightarrow e^{a\prime }(x)=  
e^b(x)\Lambda_b{}^a(x) \, ,  
\; \nonumber \\ & 
\Lambda_a{}^c \eta_{cd} 
\Lambda_b{}^d = \eta_{ab}\;  =diag(1,-1, \ldots ,-1)\,  . 
\end{eqnarray} 
This local Lorentz symmetry is the first gauge symmetry of 
gravity theory to be noted. Its infinitesimal form is 
\begin{eqnarray}
 \label{LL1} 
&  \delta_{L} x^\mu = 0 \; , \quad \delta_{L} 
e^{a}(x)=  e^b(x)L_b{}^a(x) \; ,  
\nonumber \\ & 
 L^{ab}(x)= - L^{ba}(x) \, , \qquad  
\end{eqnarray} 
where $\Lambda_b{}^a(x)= \delta_b{}^a+ L_b{}^a(x)+ {\cal O}(L^2)$. 
It is convenient to introduce the spin connection 
\begin{eqnarray}
\label{gw}
& w^{ {a} {b}}(x)= dx^{\mu} w_{\mu}^{ {a} {b}}(x)=-w^{ {b} {a}}(x)\; , 
\end{eqnarray}
with the transformation rule 
$ w^{ {a} {b}}(x) \rightarrow w^{ {a} {b}\prime}(x)= 
(\Lambda^{-1} (x) d \Lambda(x))^{ {a} {b}} + 
(\Lambda^{-1} (x) w(x)\Lambda(x))^{ {a} {b}}$, or 
\begin{eqnarray}
\label{Lvgw} & 
\delta_{L}  w^{ {a} {b}}(x)= {\cal D}L^{ab}\equiv 
dL^{ab} - 2L^{[a|c}w_c{}^{|b]}\; .  
\end{eqnarray}
The spin connection can be either expressed from the beginning through the 
vielbein field by imposing the covariant torsion constraint 
\begin{eqnarray}
& \label{gTa}  
T^a := {\cal D}e^a\equiv d{e}^{ {a}}-  {e}^{ {b}} \wedge  
 {w}_{ {b}}^{~ {a}}= 0 \; 
\end{eqnarray}
(second order approach), or treated as an independent variable in the action 
principle (first order approach). 
In both cases the dynamics is determined by the {\sl Einstein--Hilbert} 
(EH) action, which can be written in terms of differential forms 
on $M^D$ as (see {\it e.g.} \cite{rheoR,rheo} and refs. therein) 
\begin{eqnarray}\label{E-Hac}
& \qquad S_{D,G}= 
\int_{M^D} R^{ {a} {b}} \wedge e^{\wedge (D-2)}_{ {a} {b}} \; ,
\end{eqnarray}
where $e^{\wedge (D-q)}_{ {a}_1\ldots  {a}_q} \equiv {1 \over (D-q)!} 
\varepsilon_{ {a}_1\ldots  {a}_q {b}_1\ldots  {b}_{D-q}}$ 
$e^{ {b}_1} \wedge \ldots 
 \wedge e^{ {b}_{D-q}}\,$ 
and $R^{ {a} {b}}$ is the Riemann curvature,  
\begin{eqnarray}\label{gRab}
& \hspace{-0.5cm}
 R^{ {a} {b}}= dw^{ {a} {b}}- w^{ {a} {c}}\wedge
w_{ {c}}^{~ {b}} \; = \qquad {} \qquad {} \nonumber \\  
& {} \qquad = {1\over 2} dx^\nu \wedge dx^\mu R_{\mu\nu} {}^{ {a} {b}} \equiv 
{1\over 2} e^d \wedge e^c R_{cd} {}^{ {a} {b}} \; . 
\end{eqnarray}

The variation of the EH 
action (\ref{E-Hac}) reads\footnote{since 
$\delta e^{\wedge (D-2)}_{ab}= - (-1)^D e^{\wedge (D-3)}_{abc}\wedge 
\delta e^c$, 
$d (e^{\wedge (D-2)}_{ab})= - (-1)^D e^{\wedge (D-3)}_{abc}\wedge de^c$ and 
${\cal D} (e^{\wedge (D-2)}_{ab})= - (-1)^D e^{\wedge (D-3)}_{abc}\wedge T^c$.} 
\begin{eqnarray}
\label{vSGD} 
& \delta S_{D,G}= - (-1)^D \int_{M^D}
M_{(D-1)a}\wedge \delta e^a + 
\nonumber \\ & \qquad {} \qquad {} \quad  + 
\int_{M^D} e^{\wedge (D-3)}_{abc} \wedge 
T^c \wedge \delta w^{ab} \; , \;  
\end{eqnarray}
where 
\begin{eqnarray}
& \nonumber \hspace{-0.3cm} M_{(D-1){a}}:= {R}^{bc} \wedge  
{e}^{\wedge (D-3)}_{abc} \equiv  dx^{\wedge (D-1)}_{\mu} 
M_a^\mu  , \; {} \; \\ \label{gMamu}
&   M_a^\mu \propto (R_{ca}{}^{cb}- {1\over 2} \delta_a{}^b R_{cd}{}^{cd})\, 
e_b{}^\mu(x) \; , 
\end{eqnarray}
is the Einstein tensor as a $(D-1)$--form.

The $\delta e^a$ variation produces 
the free Einstein equation 
\begin{eqnarray}
\label{gM=0} & 
M_{(D-1)a}= 0 \;  \Leftrightarrow \;  
R_{ca}{}^{cb}- {1\over 2} \delta_a{}^b R_{cd}{}^{cd}=0 \; .
\end{eqnarray}
In the {\sl first order approach},  
eq. (\ref{vSGD}) 
also gives 
the torsion constraint (\ref{gTa})
\begin{eqnarray}
\label{vspin} 
& {\delta S_{D,G}\over \delta w^{ab}_\mu (x)} =0  \; \Rightarrow 
\quad e^{\wedge (D-3)}_{abc} \wedge 
T^c = 0  \nonumber \\   & \qquad \Rightarrow \quad 
T^c \equiv {1\over 2} e^b\wedge e^a T_{ab}{}^c =0\; .
\end{eqnarray}
In the {\sl second order approach}, 
the torsion constraint (\ref{gTa}) is imposed {\it ab initio}, 
so that 
\begin{eqnarray}
\label{vSGDII} & \hspace{-0.3cm} 
\delta S_{D,G}\vert_{T^a=0}= - (-1)^D \int_{M^D}
M_{(D-1)a}\wedge \delta e^a \; .
\end{eqnarray}
Thus, varying the EH action 
one can ignore the dependence of the spin connection on the vielbein
(see \cite{New81} and refs. therein).

\section{Gauge symmetries of General Relativity} 

\subsection{Diffeomorphism invariance}

All the above formalism is written in terms of differential 
forms on spacetime. 
Clearly, these are invariant under an arbitrary change of  
local coordinates {\it i.e.}, they are 
$M^D$--diffeomorphisms invariant 
\begin{eqnarray}\label{dif}
& {x}^\mu \mapsto {x}^{\mu\prime}= {x}^\mu + \delta_{diff} {x}^\mu 
\equiv x^\mu + b^\mu (x) \, , 
\\ \label{edif} 
& e^{a}(x)\mapsto  e^{\prime a}(x^{\prime})= e^{a}(x) \; , \; 
\mbox{etc.} 
 \end{eqnarray}
In field theory 
analyses, where {\sl only} fields, such as  
$e_\mu^a(x)$, are considered as 
dynamical variables, 
this obvious invariance can be ignored in favour of general coordinate 
symmetry (see below). However,  
when the coupled system of (super)gravity and (super)brane 
is considered in the framework  of an action principle (see \cite{BAIL,BAI+}) 
the set of dynamical variables includes, besides the fields 
 $e_\mu^a(x)$ {\it etc.}, the local coordinate functions 
$\hat{x}^\mu(\xi)$ defined by the map $\hat{\phi}:W^{p+1}\mapsto M^D$, 
where  $W^{p+1}$ is the 
worldvolume with local coordinates 
$\xi^i = (\tau , \sigma^1, \ldots , \sigma^p)$. 
This suggests adopting a field--space democracy approach 
\cite{fsdem} where the fields ($e_\mu^a(x)$ {\it etc.}) and the spacetime 
coordinates $x^\mu$ are treated on equal footing.

\subsection{General coordinate symmetry and its `variational version'}

Besides being local Lorentz and diffeomorphism invariant,  
the EH action (\ref{E-Hac}) is invariant under general coordinate 
transformations as we discuss below. 

To derive the equations of motion for a field theory 
from the variational principle,  
as {\it e.g.} eq. (\ref{gM=0}) for general relativity, 
one uses arbitrary variations of the fields only, {\it e.g.} 
$\delta^\prime e_\mu^a(x)$, so that 
\begin{eqnarray}\label{vprime}
&\delta^\prime x^\mu = 0\; , \qquad   
\delta^\prime e^a(x) =  dx^\mu \delta^\prime e_\mu^a(x)\; .
 \end{eqnarray}
On the other hand, a general variation $\delta e^a(x)$ is 
\begin{eqnarray}\label{vgen}
& \qquad 
\delta e^a(x) \equiv \delta^\prime e^a(x) + \delta_{\delta x} e^a(x) \; , 
 \end{eqnarray}
where $\delta_{\delta x}$ denotes the variation due to the change 
$x\rightarrow x^\prime= x+\delta x$. 
The $\delta_{\delta x}$ variation is given by the Lie derivative 
$L_{\delta x}= d i_{\delta x} + i_{\delta x} d$. For instance,    
\begin{eqnarray}\label{vdx}& 
\delta_{\delta x} e^a(x) := e^a(x+ \delta x) - e^a(x)  = \qquad 
\nonumber  \\ \nonumber 
& = L_{\delta x} e^a(x) =
d i_{\delta x}e^a(x) + i_{\delta x} de^a(x) = 
\\ 
& = 
{\cal D}  (i_{\delta x}e^a(x) ) +  i_{\delta x} T^a(x) + 
e^b \, i_{\delta x} w_b{}^a \; ,
 \end{eqnarray}
where 
\begin{eqnarray}\label{ide} 
& i_{\delta x} e^a = \delta x^\mu e_\mu^a(x)\; , 
\nonumber  
\quad  i_{\delta x} w^{ab}=  \delta x^\mu w_\mu^{ab}(x) \; , \\ 
& i_{\delta x} T^a(x) =  e^b i_{\delta x} e^cT_{cb}{}^a(x) \; ; 
 \end{eqnarray}
the last term in (\ref{vdx}), 
$e^b  i_{\delta x} w_b{}^a = 
e^b \, \delta x^\mu w_\mu{}_b{}^{a}(x)$, is the local Lorentz rotation 
induced by $\delta x^\mu$.

In the above notation,  diffeomorphism invariance 
((\ref{dif}), (\ref{edif})) can be formulated as a symmetry under the 
transformations  
\begin{eqnarray}
\label{ddif}
& \delta_{diff} {x}^\mu   =  b^\mu ({x}) \;  , 
\\ \label{eddif0} 
& \delta_{diff} e^{a}(x)= \delta^\prime_{diff} e^{a}(x) + 
L_b e^{a}(x)=0 \; . 
 \end{eqnarray}
Thus, $\delta^\prime_{diff} e^{a}_\mu(x)$ is defined by 
$- (L_b e^a)_\mu$, {\it i.e.}
\begin{eqnarray}\label{eddif} 
& \qquad \delta^\prime_{diff} e^{a}(x)= - L_b e^a \; ,
 \end{eqnarray}
and $\delta_{diff}S_{D,G}=0$ follows as an 
evident consequence of (\ref{vSGDII}) and (\ref{eddif0}).

In contrast, general coordinate transformations or local translations 
are defined as arbitrary displacements of the spacetime points, 
 \begin{eqnarray}\label{gc}
& {x}^\mu \mapsto {x}^{\mu\prime}= {x}^\mu + \delta_{gc} {x}^\mu 
\equiv x^\mu + t^\mu (x) \, 
 \end{eqnarray}
({\it cf.} (\ref{ddif})) which {\sl induce} 
differential forms transformations as {\it e.g.}
 \begin{eqnarray}
\label{gce} 
& \hspace{-0.3cm} 
e^{a}(x)\mapsto  e^{a}(x^{\prime})=  e^{a}(x+\delta_{gc} x )\equiv 
e^{a}(x+t) \; , 
\end{eqnarray}
{\it  i.e.}, 
 \begin{eqnarray}
& \delta_{gc} e^{a}(x)= \delta_{\delta_x} e^{a}(x)
\equiv \delta_{t} e^{a}(x)= 
L_t e^{a}(x)=  \nonumber \\ \label{dgce}  & \quad = 
{\cal D}  (t^a(x) ) + i_{t} T^a(x) + 
e^b  \, i_{t} w_b{}^a \;   
\end{eqnarray}
(as in  (\ref{vdx})) where 
$ t^a(x)= i_te^a(x)\equiv t^\mu (x) e_\mu^a(x)\,$.

Consider a $D$--form ${\cal L}_D$ on $M^D$, involving the 
$\wedge$ product of forms, their exterior derivatives and, 
possibly, the $*$ Hodge operator. Then   
${\cal L}_D(x)\mapsto {\cal L}_D (x^{\prime})=  
{\cal L}_D (x+t)$,  and  
 \begin{eqnarray}\label{gcLD} 
& \hspace{-0.2cm} \delta_{gc} {\cal L}_D = 
L_b {\cal L}_D =  
(i_td + di_t){\cal L}_D = d(i_t{\cal L}_D) \; ,  
\end{eqnarray}
since  any $D$--form on $M^D$ is closed. Thus, 
{\sl any} $S=\int_{M^D}{\cal L}_D$ is general coordinate invariant. 
In particular, the EH action (\ref{E-Hac}) 
possesses this invariance. 

Thus, one may look at $\delta_{diff}$ and $\delta_{gc}$, respectively, as 
passive and active forms of the same spacetime coordinates transformations, 
unaffecting  any theory defined through an integral of  
a $D$--form ${\cal L}_D$ on $M^D$. 
This picture changes when the action of a $p$--brane, which is given by 
an integral $\int_{W^{p+1}}\hat{{\cal L}}_{p+1}$ 
on the $(p+1)$--dimensional worldvolume ${W}^{p+1}$,  
is considered together with $\int_{M^D}{\cal L}_D$.
The coupled action $\int_{M^D}{\cal L}_D+ \int_{W^{p+1}}\hat{{\cal L}}_{p+1}$ 
  will still possess spacetime diffeomorphism invariance provided 
 $\hat{{\cal L}}_{p+1}$ is formulated in terms of pull--backs of  
spacetime differential forms 
(so that eq. (\ref{dif}) implies $ 
\hat{x}^\mu (\xi) \mapsto \hat{x}^{\mu\prime}(\xi)= \hat{x}^\mu(\xi) + 
b^\mu (\hat{x}(\xi))\;$), 
but it will not be spacetime general coordinate invariant, 
since such an invariance means equivalence between different 
spacetime points, and points on the brane are not equivalent to 
points outside it. 
 
Let us go back to the pure gravity case.
On account of diffeomorphism invariance, 
one can use equivalently (rather than 
$\delta_{gc} (t)$, eqs. (\ref{gc})--(\ref{dgce})), 
 $\delta_{gc} (t)$ followed by  a  diffeomorphism 
(eqs. (\ref{ddif}), (\ref{eddif}))
with $b^\mu(x) = -t^\mu(x)$,  $\delta_{diff} (b=-t)$. 
As we are dealing with  a local Lorentz invariant theory, we may also add 
a local Lorentz transformation with parameter 
$L^{ab}= - i_tw^{ab}$ and get 
  \begin{eqnarray}\label{tdgc} 
& \tilde{\delta}_{gc}(t^a) := 
{\delta}_{gc}(t^\mu) + {\delta}_{diff}(b^\mu= - t^\mu) + \nonumber \\ 
& \qquad {} \qquad + 
{\delta}_{gc}(L^{ab}= - i_tw^{ab}) \; .
\end{eqnarray}
This $\tilde{\delta}_{gc} (t^a)$  
will be called, following \cite{WZ78},
the `variational version' of the general coordinate transformation 
${\delta}_{gc}(t^a)$. 
Indeed, it does not act on $x^\mu$, 
 \begin{eqnarray}\label{ltx} 
& \quad \tilde{\delta}_{gc}x^\mu := 0 \; , \qquad 
\end{eqnarray}
so that, {\it e.g.} (eqs. (\ref{vprime}), (\ref{dgce})),  
\begin{eqnarray}\label{lt}
& \tilde{\delta}_{gc}  e^{a}(x):= 
{\cal D}t^a + e^ct^b T_{bc}{}^a(x)= 
 \tilde{\delta}^\prime_{gc}  e^{a}(x)\, . 
\end{eqnarray}
Thus  $\tilde{\delta}_{gc}$ provides the complete general coordinate 
variation of a differential form, 
{\it e.g.} $\delta_{gc} e^a(x)= \tilde{\delta}_{gc} e^a(x)$, as a result  
of the field variation $\tilde{\delta}_{gc}^\prime$ only.

In the second order approach, where $T^a=0$, 
the  $\tilde{\delta}_{gc}$ transformations (\ref{lt}) simplify   
and acquire the characteristic form of a gauge field 
transformation,   
\begin{eqnarray}
\label{ltII}
& \tilde{\delta}_{gc}  e^{a}(x)\vert_{T^a=0}= {\cal D}t^a \; .
\end{eqnarray}
This provides the possibility of treating gravity as 
a gauge theory of 
local translations 
in their variational form $\tilde{\delta}_{gc}$ 
(see \cite{GRgauge} for early discussions of gravity as a gauge theory).

Note, finally, that since the above $D$--form ${\cal L}_D$ is diffeomorphism 
invariant and $\delta_{gc}S=0$ by (\ref{gcLD}), it follows directly 
from (\ref{tdgc})
that $\tilde{\delta}_{gc}S=0$ also.

\section{Second Noether theorem applied to General Relativity} 

The invariance of the EH action 
(\ref{E-Hac}) under the variational version of the 
general coordinate transformations $\tilde{\delta}_{gc}(t^a)$, 
\begin{eqnarray}
\label{vgSGII} & 
\hspace{-0.2cm} \tilde{\delta}_{gc} S_{D,G}=  (-1)^D \int_{M^D}
{\cal D}M_{(D-1)a}\; t^a(x) = 0 \, , \;
\end{eqnarray}
follows from the fact that the Bianchi identity 
${\cal D}R^{ab}\equiv 0$ and the torsion constraint  
(\ref{gTa}) imply 
\begin{eqnarray}
\label{DM=0} & 
\qquad {\cal D}M_{(D-1)a}\equiv  0 \; 
\end{eqnarray}
(for simplicity, we use in this section the second order approach). 
This is the so--called {\sl Noether identity} (NI) which
reflects the presence of a gauge symmetry, 
here the symmetry under $\tilde{\delta}_{gc}$.  

In general, the second Noether theorem states that 
any gauge symmetries, $\delta_{gauge}S=0$, given by transformation rules 
that involve the derivatives of the local parameter up to $k$-th order, are 
in one--to--one 
correspondence with their associated NIs, {\it i.e.} with 
identically satisfied relations between the ({\it l.h.} sides of the) 
Lagrangian equations of motion and their derivatives up to $k$-th order. 

To discuss also $\delta_{diff}$ and $\delta_{gc}$ in this framework, 
consider the second order approach to $D$--dimensional general relativity 
in a field--space democracy context \cite{fsdem}, where 
coordinates 
and fields are treated on the same footing. Then, the 
dynamical variables are the vielbein field $e_\mu^a(x)$ 
and the spacetime coordinate $x^\mu$.  
Their Lagrangian equations are 
\begin{eqnarray}
\label{Nx} & {\cal N}_\mu=0 \;  ,  
\quad {\cal N}_\mu := {\delta S \over \delta x^\mu} \; 
 \qquad \mbox{and} \qquad 
 \\ \label{Ma}
& \qquad  M_a^\mu=0\; , \quad 
M_a^\mu := -(-1)^D {\delta S \over \delta e_\mu^a(x)}\; , \qquad  
\end{eqnarray}
where $M_a^\mu$ defines 
the differential $D$--form  $M_{(D-1){a}}$, eq.  
(\ref{gMamu}). 

To find an explicit expression for ${\cal N}_\mu$ one uses the splitting 
(\ref{vgen}) of the general variation of the 
EH action (\ref{vSGDII}) 
\begin{eqnarray}
\label{vSGDIIs} & 
\delta S_{D,G}= - (-1)^D \int_{M^D}
M_{(D-1)a}\wedge \qquad \nonumber \\ & \qquad {} \qquad {} \qquad 
\wedge (\delta^\prime e^a(x) + \delta_{\delta x} e^a(x))\; . 
\end{eqnarray}
The Einstein equation (\ref{Ma}) now follows from the $\delta^\prime e^a(x)$ 
variation, while the $\delta x^\mu$ variation entering  
$\delta_{\delta x} e^a(x)= 
{\cal D}  (i_{\delta x}e^a(x) ) + e^b \, i_{\delta x} w_b{}^a\,$ 
(eq. (\ref{vdx}) with $T^a=0$)
results in eq. (\ref{Nx}) with $ {\cal N}_\mu$ defined by 
\begin{eqnarray}
\label{calN=} & 
d^Dx{\cal N}_\mu= (-1)^D {\cal D}M_{(D-1)a}\, e_\mu^a - \qquad \nonumber \\ 
& \qquad {} \qquad {}  - (-1)^D
 M_{(D-1)a}\wedge e^b w_{\mu b}{}^a \; .
\end{eqnarray}

The variational version 
of the general coordinate transformations  
$\tilde{\delta}_{gc}$, (\ref{ltx})--(\ref{ltII}), as well as the          
local Lorentz transformations ${\delta}_{L}$, (\ref{LL1}), 
do not act on the spacetime coordinates. 
As a result, the NIs reflecting the invariance under $\tilde{\delta}_{gc}$ 
and ${\delta}_{L}$ only  involve 
the {\it l.h.} side of eq. (\ref{gMamu}) (eqs. (\ref{Ma})) 
\begin{eqnarray} 
\label{DM=NI} & \tilde{\delta}_{gc}S_{D,G}=0 \quad \Leftrightarrow \quad 
{\cal D}M_{(D-1)a}\equiv  0 \; \; ,  \\  
\label{LLNI} & 
\tilde{\delta}_{L}S_{D,G}=0 \quad \Leftrightarrow \quad 
M_{(D-1)[a}\wedge e_{b]} \equiv  0 \; \; .   
\end{eqnarray}  
 
In contrast, for the general coordinate symmetry in its original form 
${\delta}_{gc}$, (\ref{gc}), (\ref{gce}), the basic transformations are 
arbitrary changes of the spacetime points, eq.  (\ref{gc}). 
Thus,  
\begin{eqnarray}
\label{gcNI} & 
{\delta}_{gc}S_{D,G}=0  \qquad \Leftrightarrow \qquad 
{\cal N}_\mu := {\delta S \over \delta x^\mu}\equiv  0 \; .
\end{eqnarray}
Using eq. (\ref{calN=}) together with (\ref{DM=NI}) and (\ref{LLNI}) one finds 
that the identity (\ref{gcNI}) holds indeed. 

It might look  strange that two equivalent formulations
of the same general coordinate 
symmetry, $\tilde{\delta}_{gc}$ and ${\delta}_{gc}$, 
have different NIs. The reason is simple:
a linear combination of these 
NIs reproduces the NI for diffeomorphism invariance 
$\delta_{diff}$,  
(\ref{ddif}),  (\ref{eddif0}) (or, equivalently
(\ref{dif}),  (\ref{edif})). 
Indeed, 
the explicit form of ${\cal N}_\mu$ (\ref{calN=}) 
actually provides us with such NI
\begin{eqnarray}
\label{diffNI} & 
{\delta}_{diff}S=0 \; \Leftrightarrow \; 
d^Dx {\cal N}_\mu -  (-1)^D \times  \nonumber \\ 
& ({\cal D}M_{(D-1)a} e_\mu^a - 
 M_{(D-1)a}\wedge e^b \; w_{\mu b}{}^a)  
\equiv  0 \; .
\end{eqnarray}
As the two terms inside the brackets are identically equal to zero due to the 
NIs for $\tilde{\delta}_{gc}$ and $\delta_{L}$, (eqs. (\ref{DM=NI}) 
and (\ref{LLNI}))  the NI 
(\ref{diffNI}) implies (\ref{gcNI}) and {\it viceversa}. This 
translates the definition of $\tilde{\delta}_{gc}$  
in eq. (\ref{tdgc}) to the language of the second Noether   theorem.

\section{$D$--dimensional supergravity}

\subsection{Local supersymmetry and supergravity}

Supergravity (see {\it e.g.} \cite{New81} and refs. therein) 
is the gravity theory invariant under  
local supersymmetry $\delta_{ls}$. This is a local symmetry involving a  
fermionic (Grassmann) spinor parameter $\epsilon^{\underline{\alpha}}(x)$, 
($\; \underline{\alpha} = 1, \ldots n$ is a $D$--dimensional 
Lorentz spinor index,    
$n =$dim$(Spin(1,D-1))$. 
Hence $\delta_{ls}(\epsilon^{\underline{\alpha}})$ 
mixes the graviton field, {\it i.e.} the vielbein $e_\mu^a(x)$, 
with a fermionic field, 
the gravitino or Rarita--Schwinger field
$\psi_\mu^{\underline{\alpha}}(x)$. Specifically,  
\begin{eqnarray}
\label{lsusyemu} & \quad 
\delta_{ls}e_\mu^a(x)= -2i \psi_\mu^{\underline{\alpha}}(x)
\Gamma^a_{\underline{\alpha}\underline{\beta}}
\epsilon^{\underline{\beta}}(x)\; .
\end{eqnarray}
Using the fermionic one--form $e^{\underline{\alpha}}(x)$, 
\begin{eqnarray}
\label{epsi} & 
e^{\underline{\alpha}}(x):= dx^\mu \psi_\mu^{\underline{\alpha}}(x)= 
 e^a \psi_a^{\underline{\alpha}}(x)\; , 
\end{eqnarray}
eq. (\ref{lsusyemu}) reads 
\begin{eqnarray}
\label{lsusye} & 
\delta_{ls}e^a(x)= -2i e^{\underline{\alpha}}(x)
\Gamma^a_{\underline{\alpha}\underline{\beta}}
\epsilon^{\underline{\beta}}(x)\; , 
\end{eqnarray}

The vector--spinor gravitino field has 
the proper index structure to be the 
gauge field for local supersymmetry. Thus it is natural to assume 
that 
\begin{eqnarray}
\label{lsusyf0} & 
\delta_{ls}\psi_\mu^{\underline{\alpha}}(x)= 
{\cal D}_\mu \epsilon^{\underline{\alpha}}(x)
\; ,  
\end{eqnarray}
or, equivalently 
\begin{eqnarray}
\label{lsusyf4} & 
\delta_{ls}e^{\underline{\alpha}}(x)= 
{\cal D}\epsilon^{\underline{\alpha}}(x)\; .   
\end{eqnarray}
The guess (\ref{lsusyf0}) or (\ref{lsusyf4})
is supported by the fact that 
the linearized form $\delta^0_{ls}$ of (\ref{lsusyf4}),  
\begin{eqnarray}
\label{lsusyf4f} & 
\delta^0_{ls}e^{\underline{\alpha}}(x)= 
d \epsilon^{\underline{\alpha}}(x)\; ,  \;  
\delta^0_{ls}\psi_\mu^{\underline{\alpha}}(x)=\partial_\mu 
\epsilon^{\underline{\alpha}}(x)\; ,
\end{eqnarray}
is an evident symmetry of the free $D$--dimensional Rarita--Schwinger
(RS) action
on flat spacetime 
\begin{eqnarray}
\label{RSac} & 
S^{RS}_{D}= - {2i\over 3} \int_{R^{D}} 
d e^{\underline{\alpha}}\wedge e^{\underline{\beta}} \wedge 
(dx)^{\wedge (D-3)}_{\mu\nu\rho} 
\Gamma^{\mu\nu\rho}_{\underline{\alpha}\underline{\beta}}\,  
\nonumber \\ & \qquad \propto 
\int d^Dx \psi_\mu^{\underline{\alpha}} 
\Gamma^{\mu\nu\rho}_{\underline{\alpha}\underline{\beta}}
\partial_\nu\psi_\rho^{\underline{\beta}} \; .
\end{eqnarray}

The first candidate for a locally supersymmetric action 
is the sum of the free EH action 
(\ref{E-Hac}) and the RS action (\ref{RSac}) 
`covariantized'  with respect to the local Lorentz transformations 
 (\ref{LL1}) 
\begin{eqnarray}\label{SDSG}
& 
S^{2+3/2}_{D} = \int_{M^D} {\cal L}^{2}_D +  \int_{M^D} {\cal L}^{3/2}_D 
\; , \\ 
\label{LD2}
& {\cal L}^{2}_D = R^{ {a} {b}} \wedge e^{\wedge (D-2)}_{ {a} {b}} \; , 
\\ 
\label{LD3/2}
& {\cal L}^{3/2}_D = 
-{2i\over 3} {\cal D} e^{\underline{\alpha}} \wedge e^{\underline{\beta}}
\wedge 
e^{\wedge (D-3)}_{ {a} {b} {c}}~
{\Gamma}^{ {a} {b} {c}}_{\underline{\alpha}\underline{\beta}} \; .
\end{eqnarray}
For $D=4$, $N=1$, $S^{2+3/2}_{D}$ is indeed locally supersymmetric under 
(\ref{lsusye}), 
(\ref{lsusyf4}) and 
provides the action for the simple 
$D=4$ supergravity 
\cite{FFN} (see also \cite{New81,VS}).
\begin{eqnarray}\label{S4SG0}
& \qquad {} \qquad S_{4,SG} =S^{2+3/2}_{D=4} \; .
\end{eqnarray}

\subsection{General structure of the supergravity action 
and equations of motion}

In higher dimensions (in particular, in $D=10,11$) the supergravity 
multiplet involves a set of antisymmetric tensor gauge fields 
$C_{ \mu_1 \ldots  \mu_q}(x)$ described by differential forms 
\begin{eqnarray}
\label{sgmultg}
& C_{q}\equiv {1\over q!} dx^{ \mu_q}\wedge \ldots\wedge   
dx^{ \mu_1} C_{ \mu_1 \ldots  \mu_q}(x) \; ,
\end{eqnarray}
($C_3$ for $D=11$; $C_2$, $C_4$ and $B_2$ in $D=10$ type IIB,   
$C_1$, $C_3$ and $B_2$ in $D=10$ type IIA, {\it etc.}) and, 
in $4<D<11$, scalar fields ({\it e.g.} dilaton $\phi$ in  $D=10$ type IIA and 
IIB and axion  $C_0$ in $D=10$ type IIB) and spinors. Thus, in general 
\begin{eqnarray}\label{SSGD}
& S_{SG,D}= 
\int_{M^D} ({\cal L}^{2}_D +  {\cal L}^{3/2}_D 
+ {\cal L}_D{}^{\leq 1})\; ; 
\end{eqnarray}
${\cal L}_D{}^{\leq 1}$ includes, in particular, the 
kinetic term for the $q$--form gauge fields  
\begin{eqnarray}
\label{Ckin} & \hspace{-0.3cm} 
\propto \int d^D x \, det(e_\mu^a) \, {\cal H}_{\mu_1 \ldots\mu_{q+1}} 
{\cal H}^{\mu_1 \ldots\mu_{q+1}} + \ldots
\; ,
\end{eqnarray}
where 
\begin{eqnarray}
 \label{Hp+2}
&  {\cal H}_{q+1} := 
d{C}_{q} - 
c_1 {e}^{\underline{\alpha}} \wedge 
{e}^{\underline{\epsilon}}  \wedge  
\bar{\Gamma}^{(q-1)}{}_{\underline{\alpha}\underline{\epsilon}} \;  
\\ \label{Hcomp}
& \qquad =
{1\over (q+1)!} dx^{\mu_{q+1} }\wedge  dx^{\mu_{1} }
{\cal H}_{\mu_1 \ldots\mu_{q+1}}(x)\; , 
\\ 
\label{bG(k)}
& \bar{\Gamma}^{(k)}{}_{\underline{\alpha}\underline{\epsilon}}:= 
{1\over k!} {e}^{a_1} \wedge \ldots \wedge  
{e}^{a_k} 
\Gamma_{a_1\ldots a_p} {}_{\underline{\alpha}\underline{\epsilon}}\; , 
\end{eqnarray}
is the generalized field strength of $C_q$.

These kinetic terms can be written in a first order form 
(which is suitable for discussing the relation with superspace approach,  
see \cite{rheoR,rheo}) if one adds to every gauge $q$--form $C_q$ an 
auxiliary antisymmetric  tensor field 
$F_{a_1\ldots a_{q+1}}(x)= F_{[a_1\ldots a_{q+1}]}(x)$. 
 These fields can be used to construct the $(q+1)$--forms and 
the $(D-q-1)$--forms 
\begin{eqnarray}
\label{Fp2} 
& F_{q+1} \equiv 
{1\over (q+1)!} e^{a_{q+1}}\wedge \ldots \wedge e^{a_1} \,  
F_{a_1\ldots a_{q+1}}(x)
\; , 
\\ 
\label{FDp2} 
& {\cal F}_{D-q-1} = e^{\wedge (D-q-1)}_{a_1\ldots a_{q+1}} \,  
F^{a_1\ldots a_{q+1}}(x)\; , 
\end{eqnarray} 
which allow us to write the kinetic term(s) (\ref{Ckin}) as 
\begin{eqnarray}\label{LDC}
& \hspace{-0.3cm} {\cal L}_D{}^{\leq 1} = c
({\cal H}_{q+1} - {1\over 2} F_{q+1}) \wedge 
{\cal F}_{D-q-1} + \ldots \, , \quad
 \end{eqnarray}
where 
the terms denoted by dots do not contain $F^{a_1\ldots a_{q+1}}(x)$. 

Indeed, the variation of $F^{a_1\ldots a_{q+1}}(x)$ leads to 
the algebraic equation 
\begin{eqnarray}
\label{H=F} 
& \quad {\cal H}_{q+1} -  F_{q+1} = 0\; \; 
\end{eqnarray} 
which identifies the auxiliary field   $F^{a_1\ldots a_{q+1}}(x)$ 
with the generalized field strength of the tensor gauge field 
$C_{a_1\ldots a_q}(x)$, 
\begin{eqnarray}
\nonumber &
F_{a_1\ldots a_{q+1}}(x)= 
(q+1) \nabla_{[a_1} C_{a_2 \ldots a_{q+1}]}+ \ldots
= \\ \nonumber & \qquad =
(q+1) e_{[a_1}^{\mu_1}  
\ldots e_{a_{q+1}]}^{\mu_{q+1}} \partial_{\mu_1} C_{\mu_2 \ldots \mu_{q+1}}
+ \ldots \; ,
\end{eqnarray}
where the dots denote the terms with torsion and fermions. 
Substituting eq. (\ref{H=F}) 
into the Lagrangian form (\ref{LDC}) one arrives at 
the standard, 
second order approach, representation for the kinetic term of 
the gauge field $C_{\mu_1 \ldots \mu_{q}}(x)$, eq. (\ref{Ckin}).
On the other hand, varying (\ref{SSGD}),  (\ref{LDC}), with respect to 
the gauge field(s) $C_{\mu_1 \ldots \mu_{q}}(x)$ one finds 
\begin{eqnarray}
\label{G}
&\hspace{-0.2cm} {G}_{(D-q)}\equiv 
d ({e}^{\wedge (D-q-1)}_{a_1\ldots a_{q+1}} F^{a_1\ldots a_{q+1}})
+ \ldots=0 \; , \; 
\end{eqnarray}
which, after the use of eq. (\ref{H=F}), 
becomes the dynamical gauge field equation.  

For future reference we note that the equation 
$\delta S_{D,SG} /\delta {w}^{ab}=0$ determines  
the `improved'  constraint on the spacetime torsion $T^a$ 
({\it cf.} (\ref{gTa})),  
\begin{eqnarray}
& \label{Ta}  \qquad 
T^a + i e^{\underline{\alpha}} \wedge e^{\underline{\epsilon}} 
\Gamma^a_{\underline{\alpha}\underline{\epsilon}}=0 \; ,  
\end{eqnarray}
while $\delta {e}^{\underline{\alpha}}$ and $\delta {e}^{a}$ 
provide the differential form 
expression 
for the RS and Einstein equations of supergravity 
\begin{eqnarray} 
\label{RS}
&\hspace{-0.2cm}  {{\Psi}}_{(D-1)\underline{\alpha}}
:= {{4i\over 3}}{\cal D} 
{e}^{\underline{\epsilon}} \wedge  
{e}^{\wedge (D-3)}_{ {a} {b} {c}}
{\Gamma}^{ {a} {b} {c}}_{\underline{\epsilon}\underline{\alpha}} 
+\ldots = 0\, , \quad \\  
\label{MD-1}
& M_{(D-1)~{ {a}}}:= {R}^{bc} \wedge  
{e}^{\wedge (D-3)}_{abc} + 
\ldots =0 \; .  
\end{eqnarray}

For simplicity, we will not consider here the cases where 
the supergravity multiplet involves scalar and spinor fields. 
Thus our basic examples are $D=3,4\; and \; 11$  
supergravity.

In the above notation
a generic variation of the supergravity action reads 
\begin{eqnarray}
&\hspace{-0.4cm} \delta S_{D,SG}=- (-1)^D \int_{M^D}
M_{(D-1)a}\wedge \delta e^a + \nonumber 
\\ \nonumber 
& \hspace{-0.4cm}
+ (-1)^D \int_{M^D} \Psi_{(D-1)\underline{\alpha}} \wedge 
\delta e^{\underline{\alpha}} + \qquad \\ \nonumber 
& \hspace{-0.4cm} + \int_{M^D} (-1)^D e^{\wedge (D-3)}_{abc} \wedge 
(T^c + i e^{\underline{\gamma}}\wedge e^{\underline{\delta}} 
\Gamma^c_{\underline{\gamma}\underline{\delta}}) \wedge \delta w^{ab}  \qquad 
\\ \nonumber & \hspace{-0.4cm}
+ c \int_{M^D} ({\cal H}_{q+1}-F_{q+1}) \wedge 
e^{\wedge (D-q-1)}_{a_1\ldots a_{q+1}}   \delta F^{a_1\ldots a_{q+1}} + \qquad 
\end{eqnarray} 
\begin{eqnarray} 
\label{vSSGD} & \qquad 
+ c(-1)^{Dp}\int_{M^D}  G_{D-q} \wedge  \delta C_q \; . \qquad
\end{eqnarray}

\medskip

\subsection{Local supersymmetry, general coordinate symmetry 
and Noether identities}

The above first order form of the supergravity action 
(see \cite{rheo}), eq. 
(\ref{SSGD}) with (\ref{LD2}), (\ref{LD3/2}) and (\ref{LDC}), 
is written in terms of differential forms on $M^D$ 
(including the covariant zero--forms $F_{a_1\ldots a_{q+1}}(x)$)
and thus is invariant under $M^D$--diffeomorphisms, 
defined by eqs. (\ref{ddif}), (\ref{eddif}) and the 
analogous ones  
for  $ e^{\underline{\alpha}}(x)$ {\it etc.}

The action of $D$-dimensional supergravity 
(\ref{SSGD}), 
being a generalization of the EH general relativity action, 
eq. (\ref{E-Hac}), 
possesses  
local Lorentz invariance (\ref{LL1}) and 
general coordinate invariance under 
$\delta_{gc}$ (eqs. (\ref{gc}), (\ref{dgce}))
or, equivalently, under its variational version 
 $\tilde{\delta}_{gc}$ (eq. (\ref{tdgc})). 
Moreover, it is invariant under local supersymmetry 
transformations $\delta_{ls}
(\epsilon^{\underline{\alpha}})$,  
\begin{eqnarray}
\label{lsusyx} & \delta_{ls}x^{\mu} =0\; ,  \\ 
\label{lsusy} 
& \delta_{ls} e^a(x) = -2i e^{\underline{\alpha}}(x) 
\Gamma^a_{\underline{\alpha}\underline{\epsilon}}
\epsilon^{\underline{\epsilon}}(x)\; ,\quad  
\\  \label{lsusyf} 
& \delta_{ls}e^{\underline{\alpha}}(x) = 
{\cal D} {\epsilon}^{\underline{\alpha}}(x)+ 
{\epsilon}^{\underline{\epsilon}}(x) 
{\cal M}_1{}_{\underline{\epsilon}}{}^{\underline{\alpha}}(x)
 \; ,\quad \\  
\label{lsusyC}  
& \delta_{ls} C_{p+1}(x) = 2c_1  {e}^{\underline{\alpha}} 
\wedge 
\bar{\Gamma}^{(p)}{}_{\underline{\alpha}\underline{\epsilon}}\; 
{\epsilon}^{\underline{\epsilon}}(x) \; , 
\\  \label{lsusyw0}
& \delta_{ls} {w}^{ab}(x)= 
{\cal W}_{1\underline{\epsilon}}^{ab}(x) {\epsilon}^{\underline{\epsilon}}(x) 
 \, ,\;  \\ \label{lsusyw}
 & \delta_{ls} {F}^{a_1 ... a_{q+1}}(x) = 
S^{a_1 ... a_{q+1}}_{\underline{\epsilon}}
{\epsilon}^{\underline{\epsilon}} (x)\, ,
\end{eqnarray}
where $S^{a_1 ... a_{q+1}}_{\underline{\epsilon}}(x)$ and the one--forms 
${\cal M}_1{}_{\underline{\epsilon}}{}^{\underline{\alpha}}(x)$, 
${\cal W}_1{}_{\underline{\epsilon}}{}^{ab}(x)$ 
are constructed from the fields of the 
supergravity multiplet 
and the auxiliary fields ${F}_{a_1 ... a_{p+2}}(x)$ 
({\it cf.} (\ref{lsusye}), (\ref{lsusyf4})).

Then, the experience of Secs. 3,4 allows one to conclude 
(actually without any further calculations) that 
the general coordinate symmetry 
in its variational form  $\tilde{\delta}_{gc}$, eqs. (\ref{ltx}), (\ref{lt}), 
and the local supersymmetry $\delta_{ls}$, eqs.   
(\ref{lsusyx})--(\ref{lsusyw}), are reflected by Noether identities 
relating the {\it l.h.} sides of the {\sl field} equations only, namely 
 \begin{eqnarray}\label{DRS}
&   
{\cal D} {\Psi}_{(D-1)\underline{\alpha}} - 
2i {M}_{(D-1) a}\wedge {e}^{\underline{\epsilon}}
\Gamma^a_{\underline{\alpha}\underline{\epsilon}} + \ldots \equiv 0\; ,  \quad
\\ 
\label{DMa=}
& {\cal D} {M}_{(D-1) a}- \ldots \equiv 0 \; , 
\end{eqnarray}
where the terms denoted by dots turn out to be 
proportional to the {\it l.h.} sides of eqs. 
(\ref{H=F})--(\ref{RS}), 
but {\sl not} of the Einstein equation (\ref{MD-1}). 
For example, for $D=4$ $N=1$ supergravity the full 
NIs (\ref{DRS}), (\ref{DMa=}) read
\begin{eqnarray}\label{DRS4}
&
{\cal D} {\Psi}_{3\underline{\alpha}} -& \hspace{-0.3cm} 
2i {M}_{3 a}\wedge {e}^{\underline{\beta}}
\Gamma^a_{\underline{\alpha}\underline{\beta}} - \qquad 
\nonumber \\ 
& & \hspace{-1cm} - \Gamma_{a\underline{\alpha}\underline{\beta}}\; 
{\cal D}e^{\underline{\beta}}\wedge 
(T^a + i e^{\underline{\gamma}}\wedge e^{\underline{\delta}} 
\Gamma^a_{\underline{\gamma}\underline{\delta}})\equiv 0\; ,
\\ 
\label{DMa=4} \nonumber 
& {\cal D} {M}_{3 a} -& 
 {1\over 2} \epsilon_{abcd}  R^{bc} \wedge 
(T^d + i e^{\underline{\alpha}}\wedge e^{\underline{\beta}} 
\Gamma^d_{\underline{\alpha}\underline{\beta}})\equiv 0\; .
\end{eqnarray}

To check that $\delta_{ls}S_{D,SG}=0$ (or $\tilde{\delta}_{gc}S_{D,SG}=0$) 
implies (\ref{DRS}) (or (\ref{DMa=})) and {\sl viceversa} 
it is sufficient to insert 
(\ref{lsusyx})--(\ref{lsusyw}) (or  (\ref{ltx}), (\ref{lt})) in the general 
expression for the supergravity action variation (\ref{vSSGD}), 
\begin{eqnarray}
& \delta_{ls} S_{D,SG}=- (-1)^D \int_{M^D}
(M_{(D-1)a}\wedge \delta_{ls} e^a - 
\nonumber \\ & \quad - \Psi_{(D-1)\underline{\alpha}} \wedge 
\delta_{ls} e^{\underline{\alpha}} +\ldots )=  \;   
\nonumber 
\\ \nonumber 
& 
= - (-1)^D \int_{M^D} (- 2i {M}_{(D-1) a}\wedge {e}^{\underline{\beta}}
\Gamma^a_{\underline{\alpha}\underline{\epsilon}} + 
 \\ \label{vacNIsusy} & \quad + 
{\cal D} {\Psi}_{(D-1)\underline{\alpha}} + \ldots )  
{\epsilon}^{\underline{\alpha}} =0 . 
\end{eqnarray}
Then, one sees again ({\it cf.} Sec. 4) that the gauge 
invariance of the action and the Noether identities imply each other.

\section{Supergravity in superspace} 

The local supersymmetry $\delta_{ls}(\epsilon^{\underline{\alpha}})$, 
eqs. (\ref{lsusyx})--(\ref{lsusyw}), has a structure which resembles 
that of the variational copy of the general coordinate transformations, 
$\tilde{\delta}_{gc}(t^a)$ (eqs. (\ref{gc}), (\ref{dgce})), 
but with a fermionic 
parameter.  The similarity can be recognized also from the structure 
of the Noether identities, (\ref{DRS}), (\ref{DMa=}).
This is one more reason for the existence  
of superspace $\Sigma^{(D|n)}$ 
(originally introduced \cite{SS74} 
in connection with  global supersymmetry) 
with coordinates 
$Z^M = (x^\mu , \theta^{{\alpha}})$, 
 ${\alpha}= 1, \ldots , n$,  where, {\it e.g.} $n=2^{[D/2]}$ for 
$N=1$, $D\not= 10$ and $N=2$, $D=10$. 
The holonomic or coordinate basis for the  cotangent superspace 
is provided by $dZ^M= ( dx^\mu , d\theta^{{\alpha}})$,  
while the general unholonomic basis 
(with $Spin(1,D-1)$ indices denoted by underlined greek letters) 
is defined by the supervielbein forms 
\footnote{Such superforms, but depending on a Goldstone fermion 
$\Theta^{\underline{\alpha}}(x)$ rather than 
on the superspace coordinate $\theta^{\beta}$, were already used in 
\cite{VS}.} 
\begin{eqnarray}\label{EA}
& E^A(Z) = (E^a(Z), E^{\underline{\alpha}}(Z) ) = 
dZ^M E_M^A(Z)\, .  
\end{eqnarray}

The differential geometry of spacetime
can be  extended to superspace \cite{WZ78}. 
In particular, introducing the spin connection superform 
\begin{eqnarray}\label{wAB}
& w^{ab}= dZ^M w_M^{ab}(Z)\; , \; 
 w_{\underline{\alpha}}{}^{\underline{\beta}}\equiv 
{1\over 4} w^{ab}\Gamma_{ab}{}
_{\underline{\alpha}}{}^{\underline{\beta}}\; , \quad 
\end{eqnarray}
one can define superspace torsions and curvature 
\begin{eqnarray}
\label{TaS}  
& {T}^{ {a}} :=  {\cal D}E^a= 
 d{E}^{ {a}}-  {E}^{ {b}} \wedge  
 {w}_{ {b}}^{~ {a}}
\nonumber \\ & \qquad  := {1\over 2} E^A \wedge E^BT_{BC}{}^{a}\; , \quad \\ 
\label{TalS} 
& T^{\underline{\alpha}}:=  {\cal D}E^{\underline{\alpha}}= 
dE^{\underline{\alpha}}- 
E^{\underline{\beta}} w_{\underline{\alpha}}{}^{\underline{\beta}} 
 \nonumber \\ & \qquad := 
{1\over 2} E^A \wedge E^BT_{BC}{}^{\underline{\alpha}}
\; , \quad \\ 
\label{RabS}
& R^{ab}:=  dw^{ab}- w^{ac}\wedge w_c{}^b \qquad 
 \nonumber \\  & \qquad := 
{1\over 2} E^A \wedge E^BR_{BC}{}^{ab}
\; , \quad 
\end{eqnarray}
as well as, when the supergravity multiplet 
contains antisymmetric tensor gauge fields $C_q(x)$, the 
generalized field strengths  
\begin{eqnarray}
 \label{Hp+2S}
&  {\cal H}_{q+1} := 
d{C}_{q} - 
c_1 {E}^{\underline{\alpha}} \wedge 
{E}^{\underline{\epsilon}}  \wedge  
\bar{\Gamma}^{(q-1)}{}_{\underline{\alpha}\underline{\epsilon}} \;  
\\ \label{HcompS}
& \qquad =
{1\over (q+1)!} E^{A_{q+1} }\wedge  E^{A_{1} }
{\cal H}_{A_1 \ldots A_{q+1}}(Z)\; , 
\\
\label{bG(k)S}
& \bar{\Gamma}^{(k)}{}_{\underline{\alpha}\underline{\epsilon}}:= 
{1\over k!} {E}^{a_1} \wedge \ldots \wedge  
{E}^{a_k} 
\Gamma_{a_1\ldots a_p} {}_{\underline{\alpha}\underline{\epsilon}}\; ,
\end{eqnarray}
of the various gauge superforms  
\begin{eqnarray}\label{CqS}& 
 C_q := 
{1\over q!} E^{A_{q+1} }\wedge  \ldots \wedge E^{A_{1} }
C_{A_1 \ldots A_{q}}(Z)\;. 
\end{eqnarray}

\subsection{Local supersymmetries of supergravity in superspace}

The differential forms on $\Sigma^{(D|n)}$ are 
invariant under arbitrary changes of coordinates, 
{\it i.e.} under local superspace diffeomorphisms,  
\begin{eqnarray}
 \label{difSZ}
& Z^M \mapsto Z^{M\prime}= Z^M + b^M(Z)\; , 
\\ \label{difS}
& 
E^A(Z) \mapsto E^{A\prime}(Z^\prime)= E^A(Z)\; , \; \mbox{etc.} \; , 
\end{eqnarray}
for which ({\it cf.} (\ref{ddif}), (\ref{eddif0}))
\begin{eqnarray} \label{ddifSZ} & \delta_{sdiff}Z^M= b^M(Z) \; , \\ 
\label{ddifS} 
& \delta_{sdiff} E^A \equiv \delta^\prime_{sdiff} E^A+ 
 L_b E^A =0 \; ,\; \mbox{etc.} \; .
\end{eqnarray}
The superspace local Lorentz transformation 
${\delta}_{L}$, with  superfield parameter $L^{ab}(Z)=-L^{ba}(Z)$,  
is also a manifest symmetry of supergravity.

The superspace general coordinate transformations are defined by an arbitrary 
change of the local superspace coordinates (as in (\ref{ddifSZ})), 
\begin{eqnarray}
\qquad   \label{gcSZ} & \delta_{sgc}Z^M= t^M(Z) \; , 
\end{eqnarray}
but, in contrast with  (\ref{ddifS}), 
\begin{eqnarray}
\label{gcS} 
& \delta_{sgc} E^A = L_t E^A = {\cal D}i_t E^{A}+ i_t T^A + 
 \nonumber \\ 
& \qquad {} \qquad {} \qquad {} \qquad {} 
+ E^B i_tw_B{}^A \; , \\ 
\nonumber 
& i_tw_B{}^A= t^M w_{MB}{}^A\; , \; i_tE^A= t^M E_{M}^A\;, \quad \mbox{etc.}
\\ 
\label{wff}
& w_{B}{}^{A}= diag(w_b{}^a , 
                  w_{\underline{\beta}}{}^{\underline{\alpha}})\; . 
\end{eqnarray}
Using the superdiffeomorphism invariance, 
the variational copy $\tilde{\delta}_{sgc}$ 
\cite{WZ78} of ${\delta}_{sgc}$ 
is defined by ({\it cf.} (\ref{tdgc}))
  \begin{eqnarray} 
& \nonumber \tilde{\delta}_{sgc}(t^A) = 
{\delta}_{sgc}(t^M) + {\delta}_{sdiff}(b^M= - t^M) + \\ \nonumber 
& \qquad {} \qquad + 
{\delta}_{L}(L^{ab}= - i_tw^{ab}) \; , \\ \label{tdgcS} 
& t^A(Z)= i_tE^A(Z)\equiv t^M (Z) E_M^A(Z)\; .
\end{eqnarray}
Again $\tilde{\delta}_{sgc}$ does not act on the superspace coordinates 
$(x^\mu, \theta^\alpha)$, 
but acts on superforms as the Lie derivative ({\it cf.} (\ref{lt}))  
 \begin{eqnarray}\label{lstZ} 
& \tilde{\delta}_{sgc}Z^M := 0 \; , \qquad 
\\  \label{lst}
& \tilde{\delta}_{sgc}  E^{A}(Z)= dZ^M \tilde{\delta}_{sgc}^\prime 
E_M^{A}(Z) :=\nonumber \\ 
& \qquad {} \qquad := {\cal D}t^A + E^Ct^B T_{BC}{}^A(Z)\, ,
\\  \label{lstw}
& \tilde{\delta}_{sgc}  w^{ab}(Z):= E^Dt^C R_{CD}{}^{ab}(Z)\; .
\end{eqnarray}

In particular, the fermionic part 
$\tilde{\delta}_{sgcf}(\epsilon^{\underline{\alpha}})$
of $\tilde{\delta}_{sgc}$, 
determined by the parameter $t^A(Z)=(0, \epsilon^{\underline{\alpha}}(Z))$, 
 $\; \tilde{\delta}_{sgcf}(\epsilon^{\underline{\alpha}}(Z)) = 
\tilde{\delta}_{sgc}(0, \epsilon^{\underline{\alpha}}(Z))$, 
can be called superspace local supersymmetry. Its action is given by 
   \begin{eqnarray}\label{lsSZ} 
& \tilde{\delta}_{sgcf}Z^M := 0 \; , 
\\  \label{lss}
& \tilde{\delta}_{sgcf}  E^{a}(Z):= E^C
\epsilon^{\underline{\beta}} T_{\underline{\beta}C}{}^a(Z)\, ,
\\  \label{lsSf}
& \tilde{\delta}_{sgcf}  E^{\underline{\alpha}}(Z):= 
{\cal D}\epsilon^{\underline{\alpha}} + E^C
\epsilon^{\underline{\beta}} 
T_{\underline{\beta}C}{}^{\underline{\alpha}}(Z)\, ,
\\  \label{lsSw}
& \tilde{\delta}_{sgcf}  w^{ab}(Z):= 
 E^D \epsilon^{\underline{\gamma}} R_{\underline{\gamma} D}{}^{ab}(Z)\; ,
\end{eqnarray}
and is similar, albeit not identical, to the straightforward extension of the 
local supersymmetry transformations $\delta_{ls}$
(eqs. (\ref{lsusy})--(\ref{lsusyw})) to superspace.  
We will see below that the desired identification 
of $\tilde{\delta}_{sgcf}\vert_{\theta=0} =
\tilde{\delta}_{sgcf}(\epsilon^{\underline{\alpha}}(x, 0))$, 
 with $\delta_{ls}(\epsilon^{\underline{\alpha}}(x))$ appears when the 
superspace constraints are taken into account.

\subsection{Superspace constraints} 

The unrestricted supervielbein and spin connection contain a large amount of 
fields (mostly unwanted). 
The supergravity multiplets  can be extracted from the supervielbein 
by imposing {\sl covariant} constraints on the superspace torsions, 
curvature and the gauge superform field strengths. 
The main constraints have the form 
\begin{eqnarray}
  \label{TaS=}
& T^a   =  i E^{\underline{\alpha}} \wedge E^{\underline{\beta}} 
\Gamma^a_{\underline{\alpha}\underline{\beta}}\; , \\ 
 \label{Hp+2S=}
&  {\cal H}_{q+1}  := 
d{C}_{q} - 
c_1 {E}^{\underline{\alpha}} \wedge 
{E}^{\underline{\epsilon}}  \wedge  
\bar{\Gamma}^{(q-1)}{}_{\underline{\alpha}\underline{\epsilon}} = 
\quad \nonumber \\ 
&  =
{1\over (q+1)!} E^{a_{q+1} }\wedge \ldots \wedge E^{a_{1} }
F_{a_1 \ldots a_{q+1}}(Z)\; , 
\end{eqnarray}
and can be derived as a straightforward extension  
of the component, first order form supergravity eqs.  
(\ref{Ta}), (\ref{H=F}) 
to superspace. 
This fact is not accidental. It reflects the existence of the so--called 
group manifold or `rheonomic' approach to supergravity 
\cite{rheoR,rheo}, which provides the bridge between the 
component and superfield formalism  
(see also Sec. 2 of \cite{BAIL} for a brief review).

\subsection{Local supersymmetry of ($D=11$) supergravity constraints}

After the constraints (\ref{TaS=}), (\ref{Hp+2S=}) are taken into account, 
the fermionic general coordinate transformations $\tilde{\delta}_{sgcf}$
 simplify. In particular,   
 \begin{eqnarray}\label{lsSZ1} 
& \tilde{\delta}_{sgcf}Z^M = 0 \; , \qquad 
\\  \label{lsS1}
& \tilde{\delta}_{sgcf}  E^{a}(Z)= - 2i 
E^{\underline{\alpha}} \Gamma^a_{\underline{\alpha}\underline{\beta}}
\epsilon^{\underline{\beta}} \; , \quad \mbox{etc.} 
\end{eqnarray}
Now one can easily see that  $\tilde{\delta}_{sgcf}  E^{a}\vert_{\theta =0}$ 
becomes identical to  
$\tilde{\delta}_{ls}  e^{a}$, 
$\; \tilde{\delta}_{sgcf}  E^{a}\vert_{\theta =0}= 
\tilde{\delta}_{ls}  e^{a}$,  
after the usual identification of the supergravity forms   
 with the leading components of superforms, 
$(E^{a}, E^{\underline{\alpha}})\vert_{\theta =0, d\theta =0}=
(e^a, e^{\underline{\alpha}}$), 
{\it etc.}, is made.

To be specific, let us consider $D=11$ supergravity
\cite{CF80} 
(${a}= 0,1, \ldots , 10$ , 
$\underline{\alpha}= 1, \ldots , 32$).   
Here the superspace 
constraints (\ref{TaS=}), (\ref{Hp+2S=}),  
\begin{eqnarray}\label{11DTa=}
     &  {T}^{{{a}}} 
       = - i
       {E}^{\underline{{\gamma}}}  \wedge
       {E}^{\underline{{\beta}}}
       \Gamma^{{{a}}}_{\underline{{\gamma}}\underline{{\beta}}}\; ,
\qquad 
\\ 
\label{11DF4=}
&       {\cal H}_4 \equiv
       d{C}_3 - 
       {1\over 2} 
       {E}^{\underline{{\alpha}}}
       \wedge
       {E}^{\underline{{\beta}}}
       \wedge
       \bar{\Gamma}^{(2)}_{\underline{{\alpha}}\underline{{\beta}}} =
\\ 
\nonumber & \qquad  {} \qquad 
 = {1 \over 4!}  {E}^{{{c}}_4} \wedge \ldots
       \wedge
       {E}^{{{c}}_1} ~
 {F}_{{c}_1\ldots
 {c}_4}\;  , 
\end{eqnarray} 
imply
\begin{eqnarray}
\label{11DTal=}
 &      {T}^{\underline{{\alpha}}} 
=   {1 \over 2}
       {E}^{b}
       \wedge
       {E}^{c}
 T_{{{c}}{{b}}}^{~~\underline{{\alpha}}} - 
 {1 \over 2}
       {E}^{{{b}}}
       \wedge
       {E}^{\underline{{\beta}}} T_{{{b}}\underline{{\beta}}}
{}^{\underline{{\alpha}}}\; , \qquad 
 \\ \label{Tbff} & 
 T_{{{b}}\underline{{\beta}}}
{}^{\underline{{\alpha}}}= {i \over 9}
({F}_{{{b}}
 {{b}}_1
 {{b}}_2{{b}}_3}
 (\Gamma^{{{b}}_1
 {{b}}_2{{b}}_3})
 _{\underline{{\beta}}}^{~\underline{{\alpha}}}
 + \nonumber \\ &  + {1 \over 8}
 {F}^{{{b}}_1
 {{b}}_2
 {{b}}_3{{b}}_4}
 (\Gamma_{{{b}}{{b}}_1
 {{b}}_2{{b}}_3{{b}}_4})
 _{\underline{{\beta}}}^{~\underline{{\alpha}}})\; ,  \\ 
\nonumber
\\ 
\nonumber  
  &     {R}^{{{a}}{{b}}}
    = -2i     {E}^{\underline{{\alpha}}}
       \wedge
       {E}^{\underline{{\beta}}} 
T_{\underline{{\beta}}}{}^{[a\underline{\gamma}} 
\Gamma^{b]}_{\underline{\gamma}
\underline{{\alpha}}} + 
\\ 
\nonumber  
& + {E}^{{{c}}} \wedge   {E}^{\underline{{\alpha}}} 
( 
i T^{{a}{b}\underline{\beta}} 
 \Gamma_{{c}\underline{\beta}\underline{\alpha}} - 
2i T_{{c}}{}^{[{a}| \underline{\beta}} 
 \Gamma^{|{b}]}_{\underline{\beta}\underline{\alpha}} ) + 
\\ \label{11DRab=}
& + {1 \over 2}
       {E}^{{{d}}}
       \wedge
       {E}^{{{c}}} ~
 R_{cd}{}^{ab} \; .
 \end{eqnarray}
Using (\ref{11DTa=})--(\ref{11DRab=}), the superspace local supersymmetry 
(\ref{lsSZ})--(\ref{lsSw}) 
takes the form 
\begin{eqnarray}\label{11DlsusyZ} 
& \tilde{\delta}_{sgcf}Z^M = 0 \; ,
\\ 
\label{11Dlsusy}
& \tilde{\delta}_{sgcf} E^{{a}}   =  -2i E^{\underline{\alpha}} 
\Gamma^a_{\underline{\alpha}\underline{\beta}}
\epsilon^{\underline{\beta}}(Z)\; ,\quad  
\\  \label{11Dlsusyf} 
& \tilde{\delta}_{sgcf}E^{\underline{\alpha}}  = 
{\cal D} {\epsilon}^{\underline{\alpha}}(Z)+ 
E^{{b}} T_{\underline{\beta}{b}}{}^{\underline{\alpha}}
\; {\epsilon}^{\underline{\beta}}(Z)  \; ,\quad
\\  \label{11DlsusyC}  
& \tilde{\delta}_{sgcf} C_{3}  =  
E^{\underline{\alpha}}\wedge \bar{\Gamma}^{(2)}
_{\underline{\alpha}\underline{\beta}}\; 
{\epsilon}^{\underline{\beta}}(Z) \; , \\  
\label{11DlsusyF}   
 & \hspace{-0.3cm} \tilde{\delta}_{sgcf} {F}_{{a}_1{a}_2{a}_3{a}_{4}}  =  3! 
T_{[{a}_1{a}_2}{}^{\underline{\alpha}} 
{\Gamma}_{{a}_3{a}_4]\, 
\underline{\alpha}\underline{\beta}} {\epsilon}^{\underline{\beta}}(Z) \; , 
\\  \label{11Dlsusyw} \nonumber 
& \tilde{\delta}_{sgcf} {w}^{{a}{b}}  =  
- 4i E^{\underline{\alpha}}\;   
T_{(\underline{\alpha}}{}^{[{a}\; \underline{\gamma}}
\; 
{\Gamma}^{{b}]}{}_{\underline{\beta})\underline{\gamma}} 
\; {\epsilon}^{\underline{\beta}}(Z) + \\ 
& \quad   +  
i E^{{c}}  (T^{{a}{b}\; \underline{\alpha}}
{\Gamma}_{{c}\underline{\alpha}\underline{\beta}}  - 2 
T_{{c}}{}^{[{a}\; \underline{\alpha}}
{\Gamma}^{{b}]}{}_{\underline{\alpha}\underline{\beta}})
{\epsilon}^{\underline{\beta}} 
 \; .
\end{eqnarray}
Setting $\theta =0$, $d\theta =0$ in eqs. 
(\ref{11Dlsusy})--(\ref{11Dlsusyw}), one arrives at the 
{\sl on--shell} version of the local supersymmetry transformations 
characteristic of  the component supergravity 
action\footnote{ 
Alternatively, substituting  $\tilde{\theta}(x)$ for 
$\theta$ in 
(\ref{11Dlsusy})--(\ref{11Dlsusyw}) one obtains 
the {\sl on--shell} version of the local supersymmetry transformations 
characteristic of the group manifold or rheonomic action 
for $D$=$11$ supergravity \cite{rheo}.} {\it i.e.}, the 
actual local supersymmetry 
transformation which leaves the action invariant differs from the 
pull--backs of (\ref{11Dlsusy})--(\ref{11Dlsusyw}) to $M^D$ 
by terms which vanish on the mass shell
\footnote{ 
Note that the restoration of such terms is an involved technical problem. 
However, the  use of the second Noether theorem can simplify the proof 
of the local supersymmetry of the action, as it allows to work with equations 
of motion instead of the general variation.}.

The discussion of the previous section suggests the following  
observation. The same transformation rules for superfields (superforms),   
(\ref{11Dlsusy})--(\ref{11Dlsusyw}) appear for the original form 
of the fermionic general coordinate transformations with 
\begin{eqnarray}\label{11DgcfZ}
& {\delta}_{sgcf}Z^M = \epsilon^{\underline{\beta}}(Z) 
E_{\underline{\beta}}^M(Z) \; , \qquad \\ \nonumber & \qquad  
\Leftrightarrow \; 
\cases{i_{\delta_{sgcf}}E^a = {\delta}_{sgcf}Z^ME_M^a=0\, , \cr 
i_{\delta_{sgcf}}E^{\underline{\alpha}}= 
{\delta}_{sgcf}Z^ME_M^{\underline{\alpha}}=  
\epsilon^{\underline{\alpha}}(Z) 
\; , \cr}
\\ 
\label{11Dgcfb}
& {\delta}_{sgcf} E^{{a}}   =  -2i E^{\underline{\alpha}} 
\Gamma^a_{\underline{\alpha}\underline{\beta}}
\epsilon^{\underline{\beta}}(Z)\; ,\quad  
\\  \label{11Dgcff} 
& {\delta}_{sgcf}E^{\underline{\alpha}}  = 
{\cal D} {\epsilon}^{\underline{\alpha}}(Z)+ 
E^{{b}} T_{\underline{\beta}{b}}{}^{\underline{\alpha}}
\; {\epsilon}^{\underline{\beta}}(Z)  \; ,\quad
\\  \label{11DgcfC}  
& {\delta}_{sgcf} C_{3}  =  E^{\underline{\alpha}}\wedge \bar{\Gamma}^{(2)}
_{\underline{\alpha}\underline{\beta}}\; 
{\epsilon}^{\underline{\beta}}(Z) \; , \\  
\label{11Dgcf-F}   
 & \hspace{-0.3cm}
{\delta}_{sgcf} {F}_{a_1a_2a_3
a_{4}}  =  3! 
T_{[a_1a_2}{}^{\underline{\alpha}} 
{\Gamma}_{a_3a_4]\, 
\underline{\alpha}\underline{\beta}} {\epsilon}^{\underline{\beta}}(Z) \; , 
\\  \label{11Dgcfw} 
& \hspace{-0.3cm} {\delta}_{sgcf} {w}^{ab}  =  
- 4i E^{\underline{\alpha}}\;   
T_{(\underline{\alpha}}{}^{[a\; \underline{\gamma}}
\; 
{\Gamma}^{b]}{}_{\underline{\beta})\underline{\gamma}} 
\; {\epsilon}^{\underline{\beta}}(Z) + \nonumber \\  
& \quad   +  
i E^{c}  (T^{ab\; \underline{\alpha}}
{\Gamma}_{c\underline{\alpha}\underline{\beta}}  - 2 
T_{c}{}^{[a\; \underline{\alpha}}
{\Gamma}^{b]}{}_{\underline{\alpha}\underline{\beta}})
{\epsilon}^{\underline{\beta}} 
 \; .
\end{eqnarray}
Thus the $D=11$ superspace constraints 
are invariant under {\sl both} ${\delta}_{sgcf}$ and $\tilde{\delta}_{sgcf}$. 
This reflects the superdiffeomorphism invariance 
(\ref{difSZ})--(\ref{ddifS}) of forms.

\section{Local supersymmetry and $\kappa$--symmetry of a superbrane in 
a supergravity background}

To see why a full account of the local 
gauge symmetries in supergravity can be relevant, let us now consider the 
standard description of a super--$p$--brane moving in a supergravity background
\footnote{Such description could be regarded as the background field 
approximation to a fully dynamical description of 
supergravity---super-$p$-brane 
system based on a coupled action including both the supergravity and 
super-$p$-brane 
Lagrangians \cite{BAIL}.}.

Consider, {\it e.g.} the supermembrane (M2--brane) in the $D=11$ supergravity 
background 
defined by the constraints (\ref{11DTa=})--(\ref{11DF4=}).  Its action 
is \cite{BST}
\begin{eqnarray}\label{SM2}
& S_{11,2}= \int_{W^3} 
{1\over 3!} * \hat{E}_{{a}} \wedge \hat{E}^{{a}} - \hat{C}_{3}
(\hat{x}, \hat{\theta})\; , 
\end{eqnarray}
where 
\begin{eqnarray}\label{hEa}
&  \hat{E}^{ {a}} = d\hat{Z}^{M}(\xi ) E_{M}^{~ {a}}(\hat{Z})
= d\xi^i \partial_i \hat{Z}^{M}  E_{M}^{~ {a}}(\hat{Z}) \, , \;  
\\    \label{hCp}
& \hat{C}_{3}={1\over 3!} d\hat{Z}^{M_3}\wedge \ldots \wedge
d\hat{Z}^{M_1} C_{M_1 M_2 M_3}(\hat{Z})\, , \; 
\end{eqnarray}  
are the pull--backs 
$\hat{\phi}^*(E^a)$, $\hat{\phi}^*(C_{3})$
of the supervielbein and gauge field superforms on the 
$\Sigma^{(11|32)}$  superspace by the map 
\begin{eqnarray}\label{W3}
& 
\hat{\phi}: W^{3}\rightarrow \Sigma^{(11|32)}\, , 
\qquad  \hat{\phi}: \xi^i \mapsto 
\hat{Z}^{M}(\xi)\; ,  
\qquad  
\end{eqnarray} 
so that $\hat{Z}^{M}= \hat{Z}^{M}(\xi)$ {\it etc.}
As supergravity is treated as a background, 
the set of dynamical variables includes only the 
local supercoordinate functions 
$ \hat{Z}^{M}(\xi )= (\hat{x}^{\mu}(\xi ), \hat{\theta}^{\alpha}(\xi ))$, 
which define the worldvolume as  a surface in superspace, 
$\{ Z^M \in \Sigma^{(11|32)}\; \vert  \; Z^M= \hat{Z}^{M}(\xi)\}$. 
Hence the basic variations, $\delta \hat{Z}^{M}(\xi )$, can be recognized as 
a counterpart of superspace general coordinate transformations 
(\ref{gcSZ}), (\ref{gcS}), and can be 
split into the bosonic and fermionic parts
\begin{eqnarray}\label{bfdZ}
& \qquad i_\delta \hat{E}^{{a}}\equiv \delta \hat{Z}^{M}(\xi)
\hat{E}_M^{a}(\hat{Z}(\xi))\; , 
\nonumber \\ & 
\qquad i_\delta \hat{E}^{\underline{\alpha}}\equiv \delta \hat{Z}^{M}(\xi)
\hat{E}_M^{\underline{\alpha}}(\hat{Z}(\xi))\; .
\end{eqnarray}
Taking into account the constraints, one finds that the 
fermionic variations of the supercoordinate functions, 
$\delta_f\hat{Z}^M(\xi)$,  
defined by ({\it cf.} (\ref{11DgcfZ}))
\begin{eqnarray}\label{vfdZ}
& i_{\delta_f} \hat{E}^{{a}}
=0\; , \qquad
i_{\delta_f} \hat{E}^{\underline{\alpha}}
\not= 0\; , \nonumber \\ & \Leftrightarrow 
  \qquad \delta_f\hat{Z}^M(\xi)= 
i_{\delta_f} \hat{E}^{\underline{\alpha}}(\xi) 
E^M_{\underline{\alpha}} (\hat{Z}(\xi))\; , 
\end{eqnarray} 
lead to 
\begin{eqnarray}
\label{M2vfb}
& \quad {\delta}_{f} \hat{E}^{a}   =  
-2i \hat{E}^{\underline{\alpha}} 
\Gamma^a_{\underline{\alpha}\underline{\beta}}
i_{\delta_f}\hat{E}^{\underline{\beta}}\; ,\quad  
\\  \label{M2vfC}  
& {\delta}_{f} \hat{C}_{3}  =  
\hat{E}^{\underline{\alpha}}\wedge 
\hat{\bar{\Gamma}}{}^{(2)}_{\underline{\alpha}\underline{\beta}}\; 
i_{\delta_f}\hat{E}^{\underline{\beta}} \; 
\end{eqnarray}
({\it cf.} (\ref{11DgcfZ}), (\ref{11Dgcfb}), (\ref{11DgcfC}), 
where the r\^{o}le of $ \epsilon^{\underline{\alpha}}(Z)$ is played 
by $i_\delta \hat{E}^{\underline{\alpha}}$). 
 Hence,  
\begin{eqnarray}\label{lsSM2}
& \delta_{f} S_{11,2}  =  
\int_{W^3} {1\over 2} *\hat{E}_{a} \wedge \delta_{f}\hat{E}^a 
- \delta_{f}\hat{C}_3 = \quad  
\nonumber \\
 & =\int_{W^3} (-i  *\hat{E}_{a} \wedge \hat{E}^{\underline{\alpha}}
{\Gamma}^a_{\underline{\alpha}\underline{\epsilon}} 
- \hat{E}^{\underline{\alpha}}\wedge
\hat{\bar{{\Gamma}}}^{(2)}_{\underline{\alpha}\underline{\epsilon}})
i_{\delta_f}\hat{E}^{\underline{\epsilon}} \, 
\nonumber \\
 & = -i \int_{W^3} *\hat{E}_{a} \wedge \hat{E}^{\underline{\alpha}}\, 
({\Gamma}^a(I- \bar{\gamma}))_{\underline{\alpha}\underline{\beta}} 
i_{\delta_f}\hat{E}^{\underline{\beta}} \; ,  \quad
\end{eqnarray}
where 
\begin{eqnarray}
\label{bgamma}
& \qquad \bar{\gamma}\equiv {i \over 3!\sqrt{|g|}}\; \epsilon^{ijk}
\hat{E}_i^a \hat{E}_j^b \hat{E}_k^c  \Gamma_{abc} \; 
\end{eqnarray}
is the well known matrix satisfying 
${\hbox{tr}}\bar{\gamma}=0$, $\bar{\gamma}^2=I$,  that 
enters in the M2-brane  
{\sl $\kappa$--symmetry} projector  $ {1\over 2} (1+ \bar{\gamma})$
\cite{BST}. 
Thus, for $i_{\delta_f}\hat{E}^{\underline{\alpha}}
 = i_{\delta_\kappa}\hat{E}^{\underline{\alpha}}:= 
(1- \bar{\gamma})^{\underline{\alpha}}{}_{\underline{\beta}}
\kappa^{\underline{\beta}}(\xi)$ we find $ \delta_{\kappa} S_{11,2}=0$, 
which expresses the fundamental  
$\kappa$--symmetry of the supermembrane \cite{BST}.

We see that when computing the fermionic variation $\delta_f$ (eqs.   
(\ref{vfdZ}), (\ref{M2vfb}), (\ref{M2vfC})) of the supermembrane 
action we actually perform a superspace fermionic general coordinate 
transformation $\delta_{sgcf}$, eqs. 
(\ref{11DgcfZ})-- (\ref{11DgcfC}), pulled back to $W^3$: 
$\phi^*( \delta_{sgcf})=\delta_{f}$.   
The variation $\delta_f$ produces 
the superbrane equations of motion on ${W}^3$, 
$\hat{\Xi}_{\underline{\beta}}:= 
*\hat{E}_{a} \wedge \hat{E}^{\underline{\alpha}}\, 
({\Gamma}^a(I- \bar{\gamma}))_{\underline{\alpha}\underline{\beta}} =0$. 
Thus, the whole variation $\delta_f$ is not a local symmetry of the 
dynamical system including the superbrane 
(otherwise, the brane dynamics would be trivial in the `fermionic' 
directions). 
However, this fermionic equation becomes an identity when multiplied by 
$(1+\bar{\gamma})\;$, {\it i.e.} 
$\; \hat{\Xi}_{\underline{\beta}}(1+\bar{\gamma})\equiv 0$.
This is the Noether identity (Sec. 4, 5.3) 
for $\kappa$--symmetry. 
On the other hand, as $\phi^*( \delta_{sgcf})=\delta_{f}$, 
this means that the breaking of  $\delta_{sgcf}$  
by the supermembrane is partial and that the part of $\delta_{sgcf}$ 
preserved on ${W}^3$ 
is given by the $\kappa$--symmetry transformations.
Moreover, as the brane action possesses manifest local Lorentz 
and diffeomorphism invariances,  
we can use (\ref{tdgc}) to conclude that 
${\delta}_{f}S_{11,2}\equiv 
{\delta}_{sgcf}S_{11,2}$ is equal to 
 $\tilde{\delta}_{sgcf}S_{11,2}$. 
Hence $\tilde{\delta}_{sgcf}S_{11,2}=0$   
for the superfield supersymmetry transformations 
(\ref{11DlsusyZ})--(\ref{11DlsusyC})  
with the parameter restricted on ${W}^3$ to be of the form 
$\epsilon^{\underline{\alpha}} (\hat{Z})=
(1- \bar{\gamma})^{\underline{\alpha}}{}_{\underline{\beta}}
\kappa^{\underline{\beta}}(\xi)$,  and 
we can state that 
$\kappa$--symmetry is just the part of the local supersymmetry which 
is preserved by the brane 
action\footnote{See {\it e.g.} \cite{BKO} for  
the relation between the local supersymmetry preserved by the bosonic brane 
{\sl solutions} 
of the supergravity equations and 
the $\kappa$--symmetry of the effective superbrane actions.}.

\section{Concluding remarks}

The above considerations indicate that
\\
{\bf i)} The $\kappa$--symmetry of the superbrane in the superfield 
supergravity background is the part of the superfield local supersymmetry 
characteristic of the supergravity constraints which is not broken 
by the presence of the superbrane. 
\\ {\bf ii)} In any complete Lagrangian description of the 
supergravity---superbrane coupled system which includes the standard 
superbrane action, the local supersymmetry will be 
partially broken. Any coupled action describing both supergravity 
and the superbrane will possess not more than $1/2$ of the local 
supersymmetry characteristic of the `free' supergravity  action.   
\\ {\bf iii)} 
As the superbrane action is written in terms of pull--backs of 
superspace differential 
forms and, possibly, the   
{\sl worldvolume} Hodge star operator, 
the complete coupled action evidently possesses  
superdiffeomorphism symmetry $\delta_{sdiff}$. 
\\  {\bf iv)} 
As the superfield local supersymmetry can be equivalently considered 
as originated either from the superspace general coordinate transformations 
$\delta_{sgcf}$, (\ref{11DgcfZ})--(\ref{11Dgcfw}), 
or from their variational copy 
$\tilde{\delta}_{sgcf}$, (\ref{11DlsusyZ})--(\ref{11Dlsusyw}), 
we conclude that 
the coupled system of {\sl supergravity and bosonic p-brane} should 
possess $1/2$ of the local supersymmetry characteristic of the 
free supergravity, if the bosonic $p$-brane appears to be the  
$\hat{\theta}(\xi)=0$ `limit' of a superbrane \cite{BAI+}. 

We hope to return to these points in forth\-coming
publications. 

\section*{Acknowledgments}

This work has been partially supported by research grants from the  
DGICYT ($\#$ PB 96-0756),  Ucrainian FFR  ($\#$ 383), INTAS 
($\#$ 2000-254) and from the Junta de Castilla y Le\'on.

\end{document}